\documentclass[aip,apl,amsmath,amssymb,floatfix,preprint,letter]{revtex4-1}
\usepackage{graphicx}
\usepackage{bm}

\begin{document}

\title{Nonlinear interferometry approach to photonic sequential logic}

\author{Hideo Mabuchi}\noaffiliation
\affiliation{Edward L.\ Ginzton Laboratory, Stanford University, Stanford, CA 94305, USA}

\date{7 August 2011}

\begin{abstract}
Motivated by rapidly advancing capabilities for extensive nanoscale patterning of optical materials, I propose an approach to implementing photonic sequential logic that exploits circuit-scale phase coherence for efficient realizations of fundamental components such as a NAND-gate-with-fanout and a bistable latch. Kerr-nonlinear optical resonators are utilized in combination with interference effects to drive the binary logic. Quantum-optical input-output models are characterized numerically using design parameters that yield attojoule-scale energy separation between the latch states.
\end{abstract}

\maketitle

\noindent
%
%
Nanophotonic engineering has grown rapidly in recent years, fueled by remarkable advances in the fabrication of high quality-factor low mode-volume (high-$Q/V$) optical resonators~\cite{Taka07,Noto11}. A number of research groups have begun to demonstrate the potential of such structures for enabling ultra-low energy (sub-fJ) optical switching based on the bulk nonlinearity of optical materials such as InGaAsP~\cite{Noza10,Kuma10}. Although large-scale integration of such nonlinear-resonator devices to form complex feedforward/feedback networks remains a formidable challenge, substantial progress is being made~\cite{Coop10,Tagu11,Feng11}, making this an opportune moment to contemplate strategies for designing signal processing circuits that leverage unique physical attributes of the nanophotonic substrate.

Here I propose a high-level approach to photonic logic based on interferometry with nonlinear components, which exploits both the strong optical nonlinearities that may be obtained in high-$Q/V$ resonators and the possibility of circuit-scale optical coherence (phase stability) that may be anticipated in monolithic nanophotonic circuits. Using a simple Kerr-type Hamiltonian to describe the cavity nonlinear optics, it is possible to derive quantum optical models for fundamental components such as a NAND gate (with fan-out) and a bistable latch. Such models can be interconnected using simple circuit algebra~\cite{Goug09a,Goug09b} and are in fact compatible with a VHDL-based schematic capture workflow for complex circuit design~\cite{Teza11}. Through numerical simulation it is possible to explore the impact of quantum fluctuations on the operation of compound devices such as the bistable latch, indicating a significant need to study circuit topologies that utilize coherent feedback~\cite{Mabu08b,Mabu11a} to suppress quantum noise in classical ultra-low power photonic signal processing~\cite{Kerc11b}.

\begin{figure}[tb!]
\begin{centering}
\includegraphics[width=0.35\textwidth]{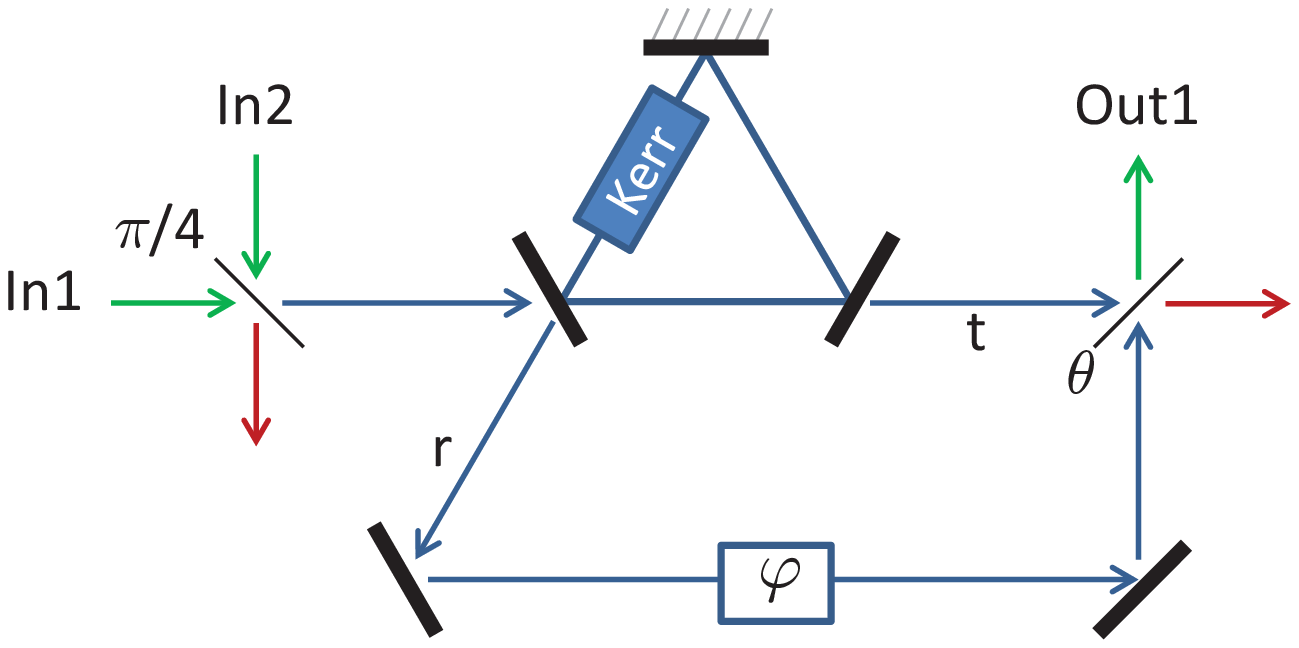}
\vbox{\vspace{0.15in}
\includegraphics[width=0.35\textwidth]{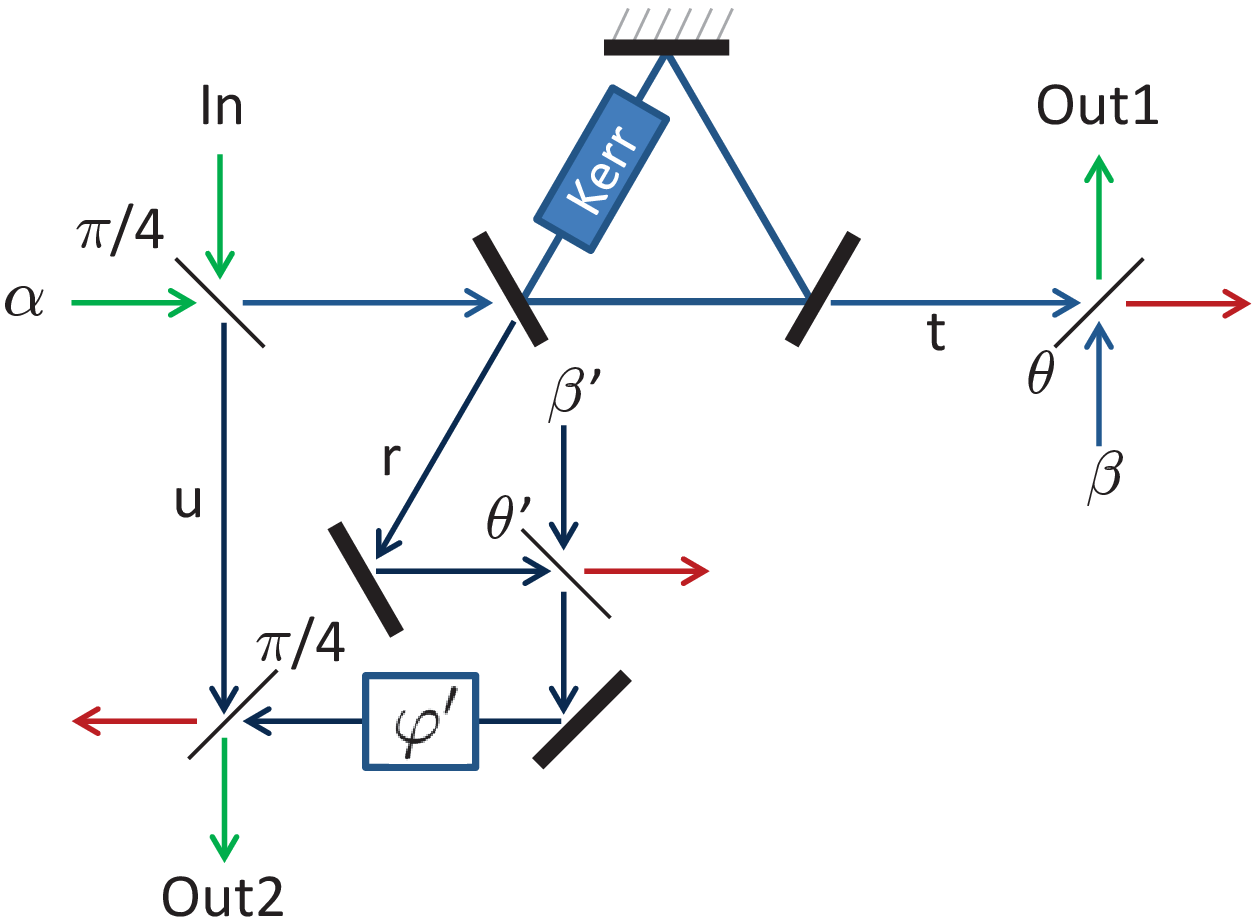}}
\end{centering}
\vspace{-0.1in}
\caption{\label{fig:gates} (UPPER) Diagram for a single-output AND gate realized by an optical interferometer incorporating a Kerr-nonlinear cavity. (LOWER) Diagram for a NOT gate with an output fan-out of two.}
\vspace{-0.1in}
\end{figure}

Signals in this logic scheme correspond to coherent states of single transverse-mode propagating electromagnetic fields, with complex amplitudes of zero and $\alpha$ representing low and high signal levels (with units such that $\vert\alpha\vert^2$ is a photon flux). The diagrams in Fig.~\ref{fig:gates} depict arrangements of optical components that realize a single-output AND gate (upper) and a NOT gate with output fanout-of-two (lower). Although discrete components are depicted, these are meant to stand for analogous arrangements of nanophotonic resonators, waveguides and splitters. The key element in each gate construction is a Kerr-nonlinear optical ring resonator (cavity) with two input-output ports. The dynamics of the internal mode of each cavity is governed by a Hamiltonian $H_0=\Delta a^\dag a + \chi a^\dag a^\dag a a,$ where we work in a rotating frame at the signal optical frequency and set $\hbar\rightarrow 1$. Here $a$ is an annihilation operator for the intracavity field and the design parameters $\Delta$ and $\chi$ represent the detuning of the cavity resonance and the Kerr nonlinear coefficient of the intracavity medium~\cite{Wall08}. The input-output coupling of the intracavity mode at port $j\in\{1,2\}$ is represented by a Lindblad operator $L_j=\sqrt{\kappa_j}a$, where $\kappa_j$ is the partial decay rate through the $j^{\rm th}$ port (the total cavity energy decay rate is $\kappa\equiv\kappa_1+\kappa_2$). The master equation for a single Kerr-nonlinear cavity is then given by
\begin{equation}
\dot{\rho}=-i[H_d,\rho]+ \sum_{j=1}^2\left\{L_j\rho L_j^\dag-\frac{1}{2}L_j^\dag L_j\rho - \frac{1}{2}\rho L_j^\dag L_j\right\},\nonumber
\end{equation}
where $H_d=H_0+i\sqrt{\kappa_1}{\cal E}(a-a^\dag)$ includes a term to represent driving the cavity through port 1 with an incident field with flux-amplitude ${\cal E}$. If $\rho_{\rm ss}$ is the steady-state solution to this master equation, then the field reflected from port 1 is $r={\cal E}+{\rm Tr}\,[\rho_{\rm ss}L_1]$ and the field transmitted through port 2 is $t={\rm Tr}\,[\rho_{\rm ss}L_2]$.

\begin{figure}[tbh!]
\begin{centering}
\includegraphics[width=0.35\textwidth]{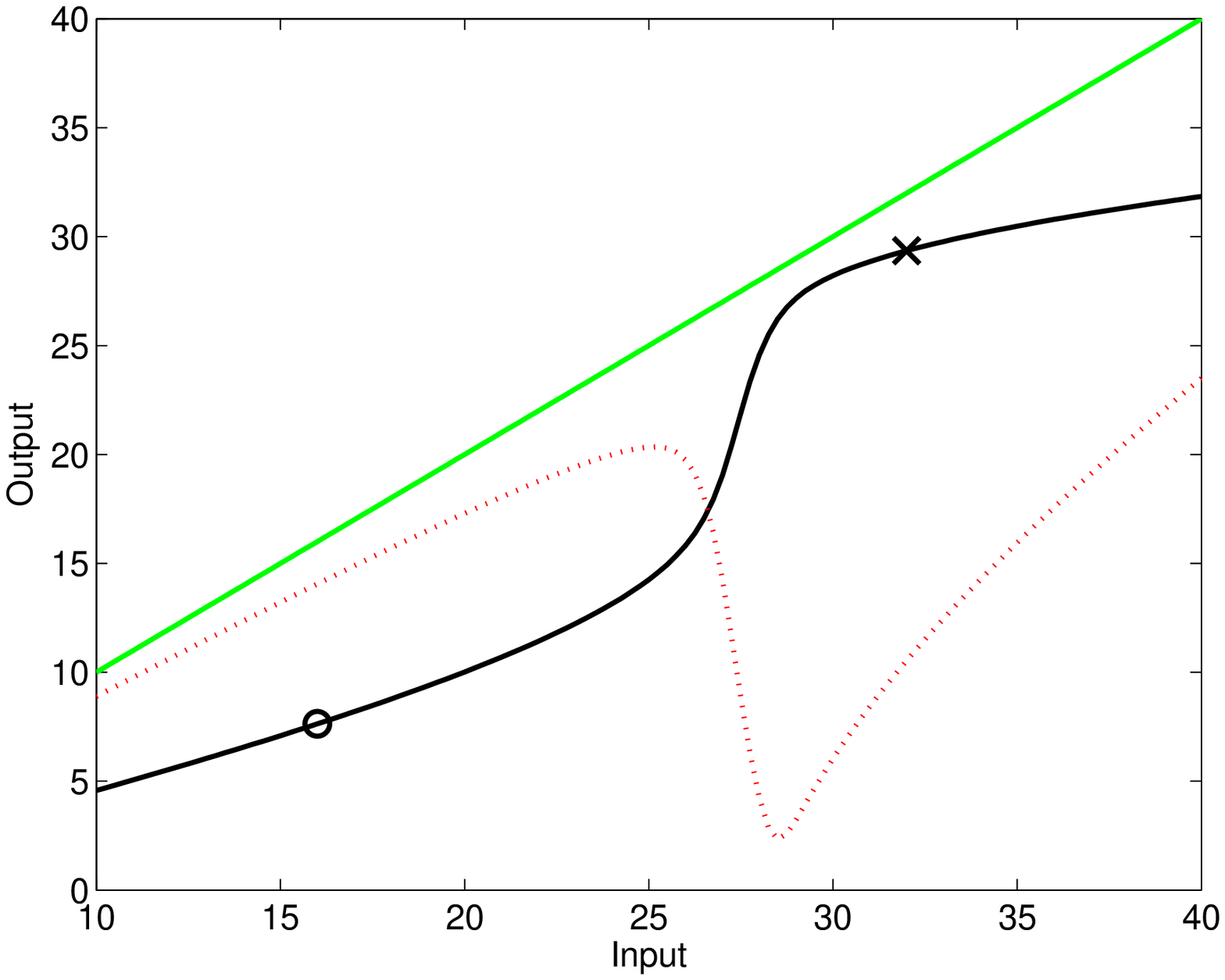}
\vbox{\vspace{0.15in}
\includegraphics[width=0.35\textwidth]{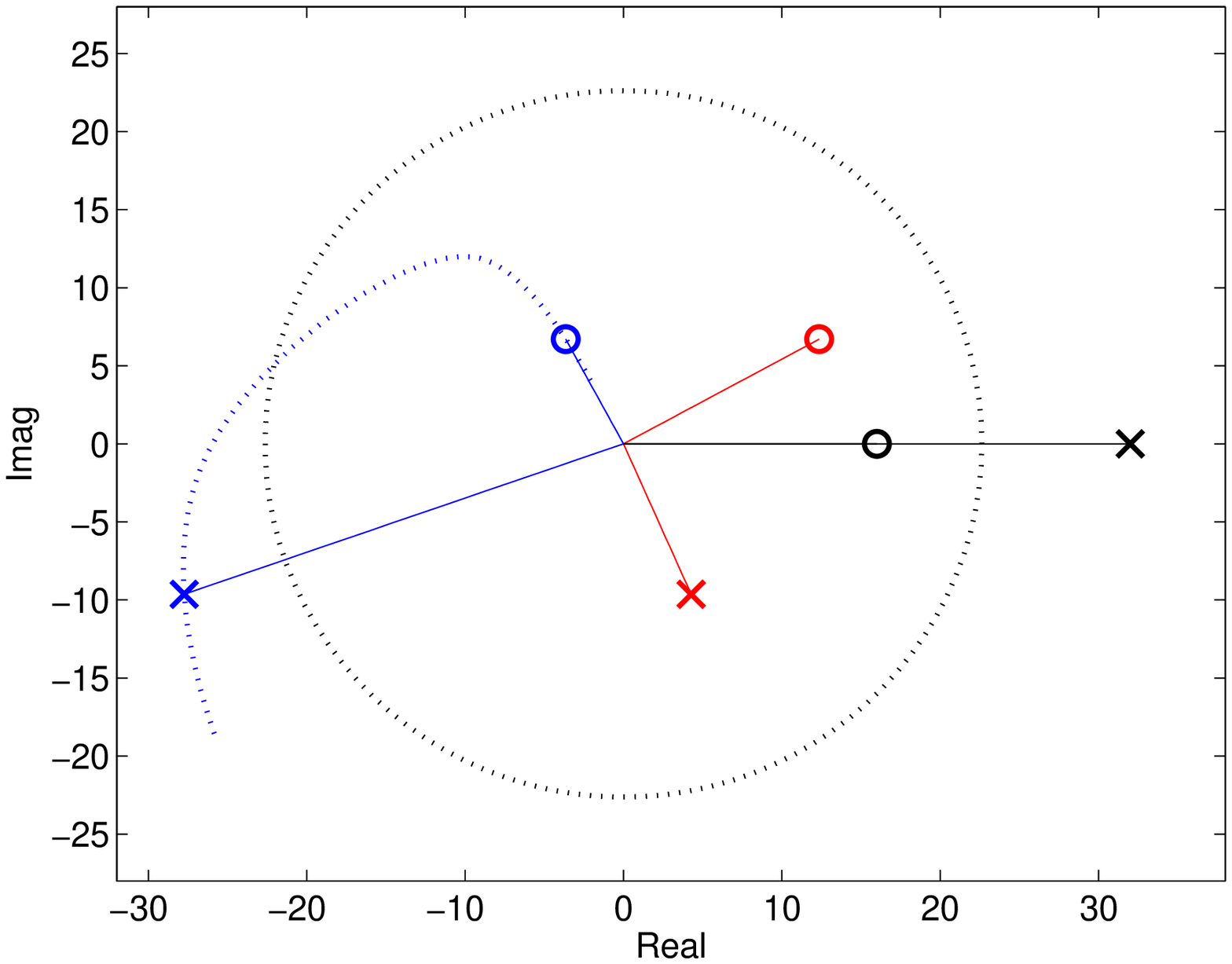}}
\end{centering}
\vspace{-0.1in}
\caption{\label{fig:Kerr} (UPPER) Steady-state flux-amplitude input-output response of a Kerr-nonlinear optical resonator (output phase shifts relative to the input field are suppressed). Solid (dotted) curve represents the cavity transmission (reflection); a solid green output$=$input reference is also drawn. (LOWER) Phasor diagram of the cavity input and output fields. Dotted black circle has radius $\alpha$. Dotted blue curve indicates the transmitted complex flux-amplitude for ${\cal E}$ in the range $10-40$. Blue (red) circles and crosses represent the steady-state transmitted (reflected) fields for inputs zero and $\alpha$, respectively. Black circle, cross represent flux-amplitudes $\alpha/\sqrt{2}$, $\sqrt{2}\alpha$.}
\vspace{-0.1in}
\end{figure}

The upper plot of Fig.~\ref{fig:Kerr} shows the steady-state response (calculated using the Quantum Optics Toolbox for Matlab~\cite{Tan99}) of a Kerr-nonlinear cavity with $\kappa_1=\kappa_2=25$, $\Delta=50$, and $\chi=-\Delta/60$. The critical ratio $\chi/\kappa$ is chosen here to obtain switching dynamics with stored energy in the range of tens of photons per cavity, which seems to represent the minimum operational energy scale that can be contemplated before quantum shot-noise fluctuations begin to dominate the device physics. In the upper plot, the horizontal axis corresponds to the input flux-amplitude ${\cal E}$ and the vertical axis corresponds to the magnitude of the flux-amplitude of the transmitted field $t$ (black solid curve) or reflected field $r$ (dotted red). Threshold behavior associated with the Kerr nonlinearity, which can be exploited to drive logic operations, is clearly visible near input $\approx 26$.

The single-output AND gate is configured with an input 50/50 beam-splitter that mixes the two input signals, such that the field driving the cavity has flux-amplitude $0$ (low+low), $\alpha/\sqrt{2}$ (low+high or high+low) or $\sqrt{2}\alpha$ (high+high). Although there is some freedom in choosing a precise value for $\alpha$, here $\alpha\rightarrow 22.6274$ in order to push $\alpha/\sqrt{2}$ well below the response threshold and $\sqrt{2}\alpha$ above. With zero driving field the overall output of the gate is clearly zero, trivially satisfying one line of the truth table. The complex amplitudes of the transmitted (blue) and reflected (red) fields are indicated in the lower plot of Fig.~\ref{fig:Kerr}, for both the $\alpha/\sqrt{2}$ (circles) and $\sqrt{2}\alpha$ (crosses) driving fields. Because of the dispersion in both the magnitudes and phases of these responses, it is possible to recombine the reflected and transmitted fields as indicated in the upper diagram of Fig.~\ref{fig:gates} to ensure that the coherent mean of the output signal $r\exp(i\varphi)\cos\theta-t\sin\theta$ is nulled for low+high and high+low inputs while the output magnitude is at least $\alpha$ for high+high inputs. For $\theta\sim1.073$, $\varphi\sim1.572$ the high+high output exceeds $\alpha$ in magnitude (adjustable by attenuation using an extra beam-splitter, not shown), with a small phase shift that can be restored to zero via propagation before reaching the next gate.

The two-output NOT gate also uses a 50/50 input beam-splitter, with constant bias $\alpha$ directed into one port as shown in the lower diagram of Fig.~\ref{fig:gates}. The field driving the cavity is then $(\alpha+{\rm In})/\sqrt{2}$ while the signal $u$ is given by $(\alpha-{\rm In})/\sqrt{2}$. Assuming ${\rm In}\in\{0,\alpha\}$ it is possible to choose values $\theta\sim0.891$, $\beta\sim-34.289-11.909i$, $\theta'\sim1.071$, $\beta'\sim7.833-17.656i$, $\varphi'\sim2.03$ such that the outputs $\beta\cos\theta-t\sin\theta$ and $((\beta'\cos\theta'-r\sin\theta')e^{i\varphi'}+u)/\sqrt{2}$ represent logical inversions of the input (plus correctable phase shifts). The AND and NOT can be cascaded to realize a universal NAND gate with fanout-of-two.

It should be noted that although the gate designs have been discussed here in terms of the coherent means of the signal fields, the cavity nonlinear dynamics do not quite preserve coherent states but rather induce small amounts of squeezing for the parameters chosen. In the ultra-low power regime (which requires strong Kerr nonlinearities) it is thus of substantial interest to study how such quantum effects propagate through a complex circuit and induce non-ideal behavior, but this is left for future studies based on more concrete implementation scenarios. Note also that Markov approximations are assumed in the cavity interactions with propagating signal fields, and that the propagation time delay between components is neglected~\cite{Goug09a} (although arbitrary propagation phase shifts can be included where desired).

\begin{figure}[tbh!]
\begin{centering}
\includegraphics[width=0.35\textwidth]{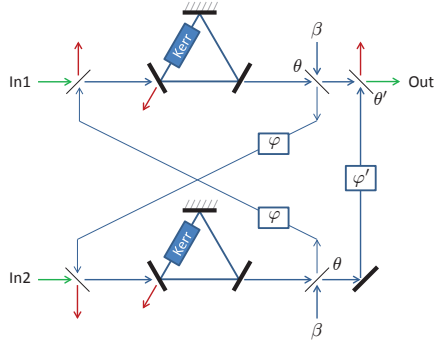}
\end{centering}
\vspace{-0.1in}
\caption{\label{fig:latchdiag} Diagram for a bistable latch realized by feedback interconnection of two pseudo-NAND gates.}
\end{figure}

\begin{figure}[tb!]
\begin{centering}
\includegraphics[width=0.35\textwidth]{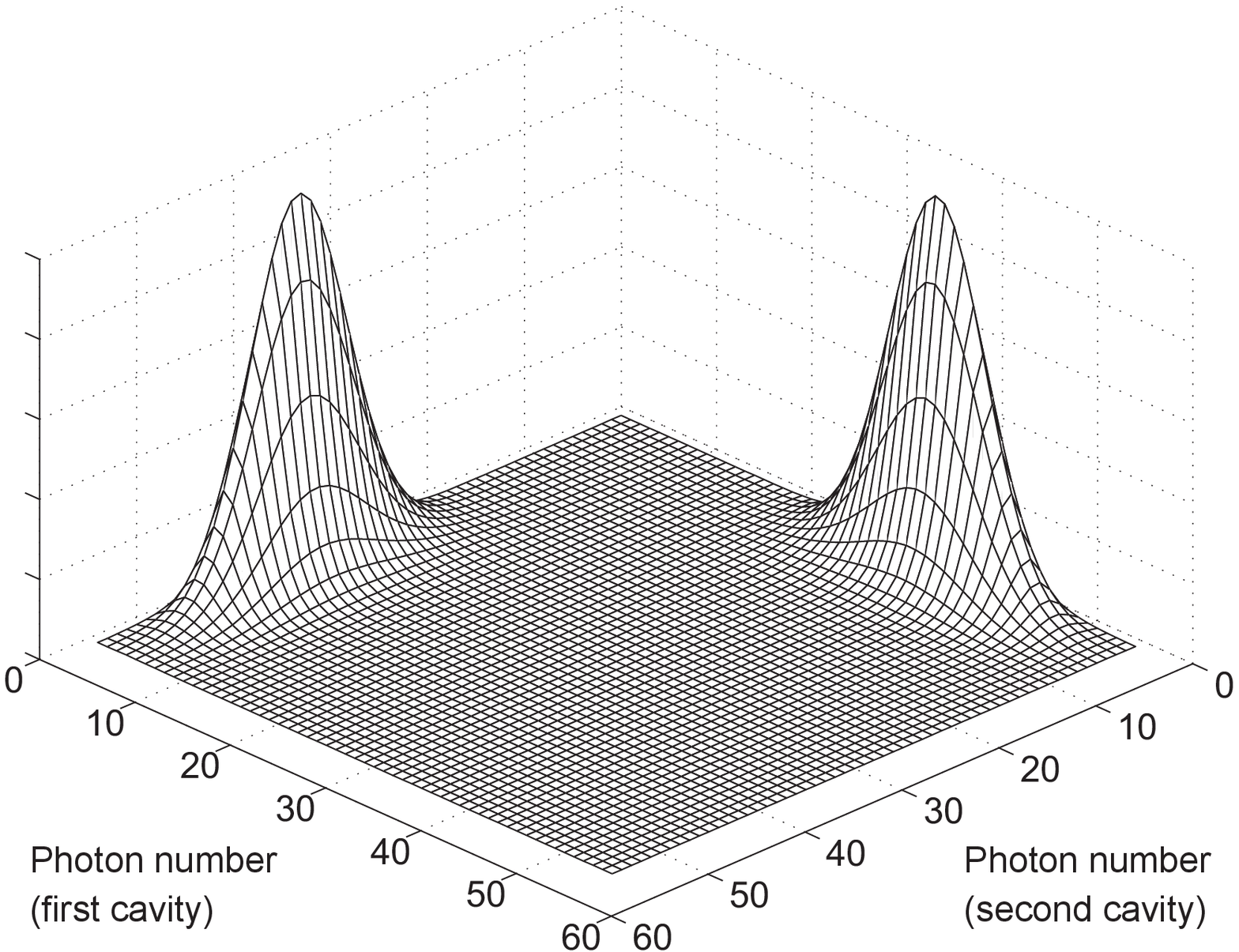}
\vbox{\vspace{0.15in}
\includegraphics[width=0.4\textwidth]{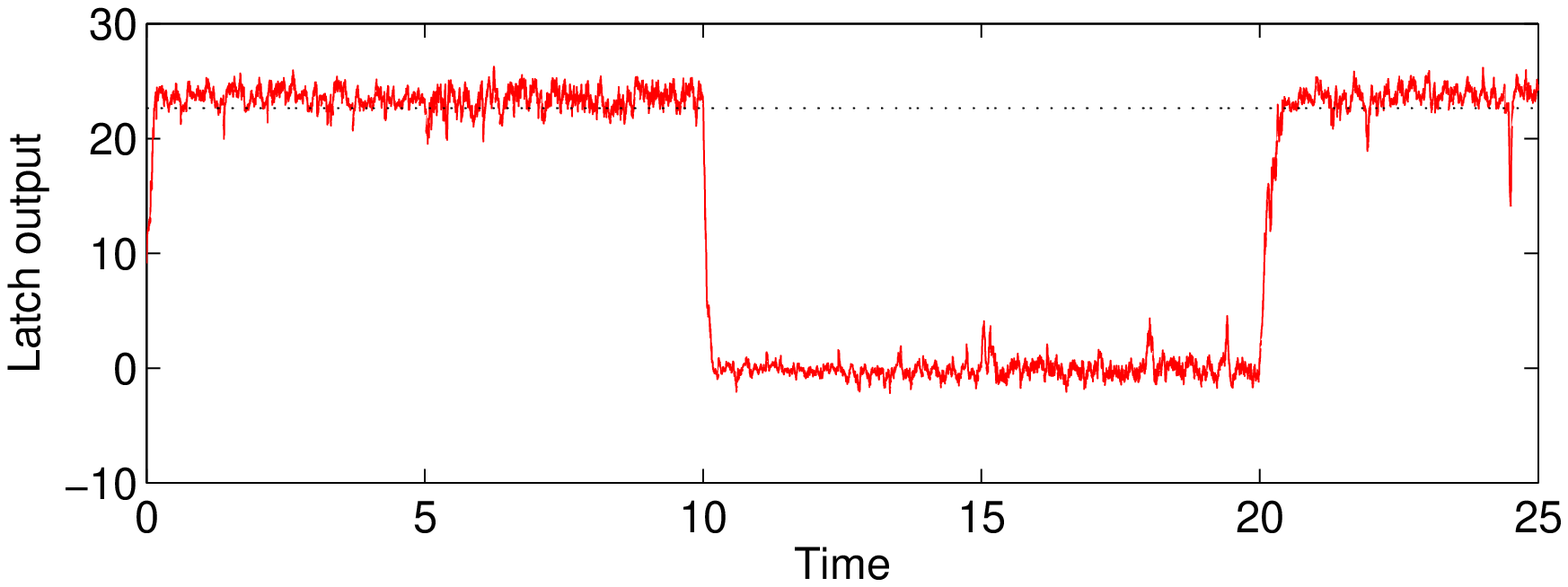}}
\end{centering}
\vspace{-0.1in}
\caption{\label{fig:latchsims} (UPPER) Surface plot of the steady-state joint photon-number distribution in the two Kerr-nonlinear cavities. (LOWER) Simulation of the latch output amplitude (corrected for an overall phase-shift) as a function of time. Initial condition is both cavities in the vacuum state and from time $0-5$ the input fields are held in the SET condition; $5-10$ HOLD; $10-15$ RESET; $15-20$ HOLD; $20-25$ SET. Dashed line is $\alpha$.}
\vspace{-0.1in}
\end{figure}

In addition to combinatorial gates, the nonlinear interferometry approach admits sequential logic components such as a bistable latch. Fig.~\ref{fig:latchdiag} depicts one possible construction based on the feedback interconnection of two Kerr-nonlinear cavities, which operates in a manner analogous to that of a canonical $\overline{S}\overline{R}$-NAND latch in electronics. With $\beta\sim-34.289-11.909i$, $\theta\sim0.891$, $\varphi\sim2.546$ the latch exhibits bistability in the photon numbers of the two cavities. A quantum-optical model corresponding to Fig.~\ref{fig:latchdiag} can be derived straightforwardly using series and concatenation products~\cite{Goug09b,Mabu11a}. As with the electronic $\overline{S}\overline{R}$-NAND latch, the HOLD condition for this device is both inputs high, while the condition with both inputs low leads to undefined response. The internal logical state of the latch (corresponding to high mean photon number in the first cavity and low mean photon number in the second, or {\it vice versa}) can be SET or RESET by pulling one input low with the other held high. The symmetric steady-state joint photon number distribution obtained with the HOLD input condition is shown in the upper panel of Fig.~\ref{fig:latchsims} and a quantum trajectory simulation of switching dynamics is shown in the lower panel ($\theta'\sim0.566$, $\varphi'\sim0.158$). Numerical analysis reveals that in this ultra-low internal energy regime, such a latch exhibits spontaneous switching due to quantum fluctuations at a rate $\Gamma\approx 0.006$, and that the switching dynamics seem to be affected by a third metastable state (in addition to the two logical states) with equal photon number in the two cavities. It would of course be desirable to optimize the latch stability and switching by varying design parameters within the space of feasible values for any given concrete implementation.

The gate and latch designs presented above serve to illustrate the basic principles of a nonlinear interferometry approach to photonic logic, exploiting the complex input-output response of Kerr-nonlinear cavities (including both phase dispersion and thresholding behavior in the signal magnitudes to realize the required truth tables) as well as the circuit-scale phase stability that may be anticipated in monolithic nano-patterned structures. Given the importance of power minimization for future information processing technologies, it seems natural to work with quantum optical models for precise quantitative analysis of the impact of quantum fluctuations on stability and dynamic behavior in nanophotonic circuits operating in the few-photon regime. Design studies that explore the performance, efficiency and robustness of candidate architectural approaches may provide crucial guidance for the continuing development of nanophotonic fabrication techniques and optical thin-film materials.

This work was supported by DARPA-MTO under grants FA8650-10-1-7007 and N66001-11-1-4106.

\end{document}